\documentclass[12pt,letterpaper]{article}
\usepackage{fullpage}
\usepackage{amsmath,amsthm,amsfonts,amssymb,amscd}
\usepackage{lastpage}
\usepackage{enumerate}
\usepackage{enumitem}
\usepackage{fancyhdr}
\usepackage{mathrsfs}
\usepackage{xcolor}
\usepackage{graphicx}
\usepackage{listings}
\usepackage[hidelinks]{hyperref}
\usepackage{algorithm2e}
\usepackage{algorithmic}
\usepackage{float}
\usepackage{tabularx}
\usepackage{booktabs}
\usepackage{subcaption}
\usepackage[colorinlistoftodos]{todonotes}
\usepackage{gensymb}
\usepackage{bm}
\usepackage{fixmath}
\usepackage{multirow}
\usepackage{ragged2e}
\usepackage{cleveref}

\title{\textbf{Optimal transmission switching and grid reconfiguration for transmission systems via convex relaxations}}
\author{Vineet Jagadeesan Nair (jvineet9@mit.edu)}

\begin{document}
\maketitle

\section{Introduction and background}

In this paper, I formulate optimization problems to perform optimal transmission switching (OTS) in order to operate power transmission grids most efficiently. In any given electrical network, several of the transmission lines are generally equipped with switches, circuit breakers, and/or reclosers. The conventional practice is to operate the grid using a static or fixed configuration. However, it may be beneficial to dynamically reconfigure the grid through switching actions in order to respond to real-time demand and supply conditions. This has the potential to help reduce costs and improve efficiency. Furthermore, such OTS may be more crucial in future power grids with much higher penetrations of renewable energy sources, which introduce more variability and intermittency in generation. Similarly, OTS can potentially help mitigate the effects of unpredictable demand fluctuations (e.g. due to extreme weather). In this project, I explored and compared several different formulations for the OTS problems in terms of computational performance and optimality. I also applied them to small transmission test case networks as a proof of concept to see what the effects of applying OTS are.

\section{Optimal power flow formulations for OTS}

Optimal power flow is the key optimization problem that governs the operations of power systems. It describes the physical laws the system needs to obey such as Kirchoff's laws, Ohm's law, line thermal limits, etc. These are exactly specified by the alternating current (AC) optimal power flow equations, shown below.

\begin{table}[]
\begin{tabular}{@{}ll@{}}
\toprule
\multicolumn{1}{c}{\textbf{Variable or parameter}} & \multicolumn{1}{c}{\textbf{Definition}}                              \\ \midrule
$n$                                            & Number of buses or nodes in network \\
$(i,j)$                                            & Bus indices                                                          \\
$L$                                                & Set of all transmission line indices                                     \\
$\mathcal{G}$     & Set of all generators \\
$a_{ij}$                                           & Binary variable indicating whether line $ij$ is open or closed       \\
$\theta_i$                                         & Voltage angle at bus $i$                                             \\
$\theta_{ij} = \theta_i - \theta_j$                & Voltage angle difference between buses $i$ and $j$                   \\
$V_i$                                              & Complex voltage at bus $i$                                         \\
$|V_i|$                                              & Voltage magnitude at bus $i$   \\
$c_{ij}$                                           & $= |V_i||V_j|cos(\theta_{ij})$                                       \\
$s_{ij}$                                           & $= |V_i||V_j|sin(\theta_{ij})$                                        \\
$e_i$                                              & $e_i = V_i^2$                                                        \\
$Y_{ij}$                                    & Admittance matrix element of line $ij$        \\
$G_{ij},B_{ij}$                                    & Real/reactive parts of admittance matrix element of line $ij$        \\
$G_{ii},B_{ii}$                                    & Real/reactive parts of admittance matrix diagonal element of bus $i$ \\
$b_{ij}$                                           & Shunt susceptance of line $ij$                                       \\
$\overline{f_{ij}}$                                & Apparent power flow capacity of line $ij$                            \\ \bottomrule
\end{tabular}
\caption{Definitions of important variables and parameters. \label{tab:def}}
\end{table}

\subsection{AC optimal power flow}

\begin{align}
     P_i+\mathbf{j} Q_i & =V_i \sum_{k=1}^n \overline{\mathbf{Y}}_{i k} \bar{V}_k 
\end{align}

Splitting these into real and imaginary components, we get:

\begin{align}
& P_i=\left|V_i\right| \sum_{j=1}^n\left|V_j\right|\left(\mathbf{G}_{i j} \cos \left(\theta_i-\theta_j\right)+\mathbf{B}_{i j} \sin \left(\theta_i-\theta_j\right)\right) \\
& Q_i=\left|V_i\right| \sum_{j=1}^n\left|V_j\right|\left(\mathbf{G}_{i j} \sin \left(\theta_i-\theta_j\right)-\mathbf{B}_{i j} \cos \left(\theta_i-\theta_j\right)\right)
\end{align}

These are clearly nonconvex constraints. A commonly used approach for transmission systems is to instead make several simplifying assumptions that result in the linear direct current (DC) optimal power flow equations instead. 

\subsection{DC optimal power flow}

\begin{align}
& \min _{P_G, Q_G,|V|, \theta} \sum_{i \in \mathcal{G}} f_i\left(P_{G i}\right) \\
& \text { subject to } \\
& P_{G i}-P_{D i}=\sum_{k=1}^n B_{i k}\left(\theta_i-\theta_k\right), \forall i \in N  \label{eq:dc_pbalance}\\
& P_{G i}^{\min } \leq P_{G i} \leq P_{G i}^{\max }, \forall i \in G \label{eq:dc_gen_caps} \\
& -\overline{f_{i k}} \leq B_{i k}\left(\theta_i-\theta_k\right) \leq \overline{f_{i k}}, \forall(i, k) \in L \label{eq:dc_flow_lims}\\
& \theta_1=0 \quad \text{(Slack bus)} \\
& \left|\theta_i-\theta_k\right| \leq \overline{\Delta \theta_{i k}}, \forall(i, k) \in L
\end{align}

These only account for power balance (\cref{eq:dc_pbalance}), generator capacities (\cref{eq:dc_gen_caps}), line power flow limits or thermal ratings (\cref{eq:dc_flow_lims}) and voltage angle constraints. While this is a reasonable approximation for high-voltage transmission grids with low line resistances, it has some key disadvantages since it completely ignores voltage magnitudes and reactive power. It also relies on a small-angle approximation to linearize all the trigonometric terms. Thus, it cannot be guaranteed that the DC-OPF solutions are feasible for the real ACOPF problem. Furthermore, the DC approximation is even worse for lower voltage distribution grids that have more losses (even though I'm restricting myself to transmission systems in this project). 

\subsection{DC OPF-based OTS}

Most of the studies I found in the literature have relied on an OTS problem formulation using DC OPF \cite{Hedman2009OptimalAnalysis,Fisher2008OptimalSwitching}, resulting in modifying the power balance constraints in \cref{eq:dc_pbalance} and line flow limits in \cref{eq:dc_flow_lims} accordingly to indicate whether a line is open or closed:

\begin{align}
& P_{G i}-P_{D i} - \sum_{k=1}^n \left( B_{i k}\left(\theta_i-\theta_k\right) +\left(1-z_{ik}\right) M \right) \geq 0,  \forall i \in N  \label{eq:dc_pbalance1_ots}\\
& P_{G i}-P_{D i} - \sum_{k=1}^n \left( B_{i k}\left(\theta_i-\theta_k\right) -\left(1-z_{ik}\right) M \right) \leq 0 ,  \forall i \in N  \label{eq:dc_pbalance2_ots} \\
& -z_{ik}\overline{f_{i k}} \leq B_{i k}\left(\theta_i-\theta_k\right) \leq z_{ik}\overline{f_{i k}}, \forall(i, k) \in L \label{eq:dc_flow_lims_ots}\\
& \sum_k\left(1-z_k\right) \leq Z,\quad z_{ik} \in \{0,1\}
\end{align}

Here the binary variable $z$ indicates whether a line is open ($z_{ik} = 0$) and no power flows through it, or closed ($z_{ik} = 1$) and all the constraints apply to it.

\subsection{Towards ACOPF-based OTS formulations}

We can reformulate the AC OPF problem by introducing additional auxiliary variables to represent the nonlinear terms (see \cref{tab:def} for details), and transform them into the following \cite{Kocuk2016StrongProblem,Wu2018RobustContingencies}:

\begin{align}
\min_{p,q,c,s,\theta} & \sum_{i \in I} \sum_{i \in G} f_i\left(P_{G i}\right) \\
& \underline{p}_i^{\mathrm{G}} \leq p_i^{\mathrm{G}} \leq \bar{p}_i^{\mathrm{G}}, \quad i=1,2, \cdots, n \\
& \underline{q}_i^{\mathrm{G}} \leq q_i^{\mathrm{G}} \leq \bar{q}_i^{\mathrm{G}}, \quad i=1,2, \cdots, n\\
& \underline{V}_i^2 \leq e_i \leq \bar{V}_i^2, \quad i=1,2, \cdots, n \\
& p_i^{\mathrm{G}}+p_i^{\mathrm{U}}-p_i^{\mathrm{D}}=G_{i i} e_i+\sum_{j=1, j \neq i}^n\left(G_{i j} c_{i j}+B_{i j} s_{i j}\right), \quad i=1,2, \cdots, n  \\
& q_i^{\mathrm{G}}+q_i^{\mathrm{U}}-q_i^{\mathrm{D}}=-B_{i i} e_i-\sum_{j=1, j \neq i}^n \left(B_{i j} c_{i j}-G_{i j} s_{i j}\right), \quad i=1,2, \cdots, n \\
% & \left(-G_{i j} e_i+G_{i j} c_{i j}+B_{i j} s_{i j}\right)^2+\left(\left(B_{i j}-b_{i j} / 2\right) e_i -B_{i j} c_{i j}+G_{i j} s_{i j}\right)^2 \leq \left(f_{i j}^{\mathrm{max}}\right)^2 \forall (i, j) \in L \label{eq:lineflow} \\
& \left(-G_{i j} e_i+G_{i j} c_{i j}+B_{i j} s_{i j}\right)^2+\left(B_{i j}e_i - B_{i j} c_{i j}+G_{i j} s_{i j}\right)^2 \leq \left(\overline{f}_{i j}\right)^2 \forall (i, j) \in L \label{eq:lineflow} \\
& c_{i j}^2+s_{i j}^2 = e_i e_j,  \quad (i, j) \in L \label{eq:voltagedef} \\
& \theta_i-\theta_j=\arctan \left(s_{i j} / c_{i j}\right) \forall(i, j) \in L \label{eq:angles} \\
& \underline {\Delta \theta} \leq \theta_i - \theta_j \leq \overline{\Delta \theta}
\end{align}

We notice that now most of the constraints are now actually convex, except for \cref{eq:voltagedef} and \cref{eq:angles}. Note that \cref{eq:lineflow} is already a convex quadratic second order conic constraint. For the first constraint \cref{eq:voltagedef}, we first relax the equality to inequality:
$$
c_{i j}^2+s_{i j}^2 \leq e_i e_j \quad \forall(i, j) \in L
$$
and then this can be transformed into a convex quadratic second-order cone programming (SOCP) constraint by defining additional auxiliary variables $D_{ij}^{1-4}$ in order to represent the bilinear right-hand side term. Thus, we can replace \cref{eq:lineflow,eq:voltagedef} with the following set of constraints:
\begin{align}
& p_{i j}=-G_{i j} e_i+G_{i j} c_{i j}+B_{i j} s_{i j}, \quad \forall(i, j) \in L \\
& q_{i j}=\left(B_{i j}-b_{i j} / 2\right) e_i-B_{i j} c_{i j}+G_{i j} s_{i j}, \quad \forall(i, j) \in L \\
& S_{i j}=\overline{f_{i j}}, \quad \forall(i, j) \in L \\
& \left(p_{i j}\right)^2+\left(q_{i j}\right)^2 \leq\left(S_{i j}\right)^2, \quad \forall(i, j) \in L \\
& D_{i j}^1=2 c_{i j}, \quad \forall(i, j) \in L \\
& D_{i j}^2=2 s_{i j}, \quad \forall(i, j) \in L\\
& D_{i j}^3=e_i-e_j, \quad \forall(i, j) \in L \\
& D_{i j}^4=e_i+e_j, \quad \forall(i, j) \in L \\
& \left(D_{i j}^1\right)^2+\left(D_{i j}^2\right)^2+\left(D_{i j}^3\right)^2 \leq\left(D_{i j}^4\right)^2, \quad \forall(i, j) \in L
\end{align}

I chose to employ a second-order conic relaxation (SOCP) here as opposed to a semidefinite programming relaxation (SDP). Although SDP relaxations have been proven to be exact for certain types of radial networks, they tend to perform poorly in practice and worse than SOCP especially for larger problems \cite{Molzahn2017ASystems,Molzahn2019AEquations}. We use this formulation for the OTS problem by introducing binary decision variables for which transmission lines or branches to keep open. While a few other papers have considered applying ACOPF models for OTS, they also rely on quite restrictive assumptions and approximations as part of their final approach \cite{Bai2017ATechnique,Hedman2011AOptimization,Jabarnejad2021ASwitching}. This motivates my goal behind this project - to apply as accurate of an OPF formulation as possible for the OTS problem.

\subsection{Optimal transmission switching problem}

\begin{align}
\min_{p,q,c,s,\theta} & \sum_{i \in I} \sum_{i \in G} f_i\left(P_{G i}\right) \\
& \underline{p}_i^{\mathrm{G}} \leq p_i^{\mathrm{G}} \leq \bar{p}_i^{\mathrm{G}}, \quad i=1,2, \cdots, n \\
& \underline{q}_i^{\mathrm{G}} \leq q_i^{\mathrm{G}} \leq \bar{q}_i^{\mathrm{G}}, \quad i=1,2, \cdots, n\\
& \underline{V}_i^2 \leq e_i \leq \bar{V}_i^2, \quad i=1,2, \cdots, n \\
& p_i^{\mathrm{G}}-p_i^{\mathrm{D}}=\sum_{j=1, j \neq i}^n p_{i j}, \quad i=1,2, \cdots, n \\
& q_i^{\mathrm{G}}-q_i^{\mathrm{D}}=\sum_{j=1, j \neq i}^n q_{i j}, \quad i=1,2, \cdots, n \\
& p_{i j}=a_{ij}\left(-G_{i j} e_i+G_{i j} c_{i j}+B_{i j} s_{i j}\right), \quad \forall(i, j) \in L \label{eq:kirchoff1}\\
& q_{i j}=a_{ij}\left(\left(B_{i j}-b_{i j} / 2\right) e_i-B_{i j} c_{i j}+G_{i j} s_{i j}\right), \quad \forall(i, j) \in L \label{eq:kirchoff2}\\
% & c_{i j}^2+s_{i j}^2 \leq e_i e_j,  \quad (i, j) \in L \; \text{(this may be redundant)} \\
& \theta_i-\theta_j=\arctan \left(s_{i j} / c_{i j}\right), \quad \forall(i, j) \in L \label{eq:angles} \\
% & S_{i j}=f_{i j}^{\max}, \quad \forall(i, j) \in L \\
& \left(p_{i j}\right)^2+\left(q_{i j}\right)^2 \leq a_{ij} \left(\overline{f}_{i j}\right)^2, \quad \forall(i, j) \in L \label{eq:flow_lim_socp}\\
& D_{i j}^1=2 c_{i j}, \quad \forall(i, j) \in L \\
& D_{i j}^2=2 s_{i j}, \quad \forall(i, j) \in L\\
& D_{i j}^3=e_i-e_j, \quad \forall(i, j) \in L \\
& D_{i j}^4=e_i+e_j, \quad \forall(i, j) \in L \\
& \left(D_{i j}^1\right)^2+\left(D_{i j}^2\right)^2+\left(D_{i j}^3\right)^2 \leq\left(D_{i j}^4\right)^2, \quad \forall(i, j) \in L \\
& a_{ij} \in \{0,1\} \quad \forall (i, j) \in L \\
& \underline {\Delta \theta} \leq \theta_i - \theta_j \leq \overline{\Delta \theta} \\
& p,q,\theta \in \mathbb{R}^n; \; c,s \in \mathbb{R}^{n \times n}; \; a \in \mathbb{R}^{|L|}
% & 0 \leq p_{ij} \leq a_{ij} S_{ij}, \quad a_{ij} \in \{0,1\}
\end{align}

This is a mixed integer nonlinear programming problem (MINLP), which is generally challenging to solve especially as we consider large network sizes with many buses/nodes and branches. We thus focus on reformulating the problematic constraints. The nonconvex constraint \cref{eq:angles} is needed to uniquely define voltage angles. Most papers I found simply discard this constraint entirely - while this is a valid approach for radial networks, it cannot be ignored for more general meshed systems. In order to convexify this, we first rewrite this constraint as follows:

\begin{align*}
    \theta_{ij} = \theta_i - \theta_j = \tan^{-1}\left(s_{i j} / c_{i j}\right) \implies \tan(\theta_{ij}) = \left(s_{i j} / c_{i j}\right)
\end{align*}

For stable operation of power systems, the voltage angle difference between neighboring buses needs to be kept small, e.g. $-1.047 = \underline{\Delta \theta} \leq \theta_{ij} \leq \overline{\Delta \theta} = 1.047 \; rad$. The bus voltage angle differences are small. Thus, we can apply a small angle approximation for $\tan()$ to relax this constraint:
$$
\lim_{\theta \to 0} \frac{\tan(\theta)}{\theta} = 1
$$

Thus, the nonconvex constraint \cref{eq:angles} can be reduced to the following bilinear equality constraint:
$$
\theta_i - \theta_j = \frac{s_{ij}}{c_{ij}} \implies s_{ij} = c_{ij} \theta_{ij} 
$$

This can further be relaxed by employing McCormick envelope (MCE) convex relaxations, replacing this bilinear term with a series of linear inequality constraints. We know
\begin{align*}
    & -|\overline{V_i}| |\overline{V_j}| \leq s_{ij} =|V_i||V_j|sin(\theta_{ij}) \leq |\overline{V_i}| |\overline{V_j}|  \\
    & -|\overline{V_i}| |\overline{V_j}|  \leq c_{ij} =|V_i||V_j|cos(\theta_{ij}) \leq |\overline{V_i}| |\overline{V_j}| \\
    & \because -1 \leq sin(\theta_{ij}), cos(\theta_{ij}) \leq 1
\end{align*}
 Generally, we aim to maintain all bus voltages within a tight interval as close to 1 p.u. possible e.g. $0.9 = |\underline{V}| \; p.u. \leq |V_i| \leq |\overline{V}| = 1.1 \; p.u.$ This allows us to bound both $c_{ij}, s_{ij} \in \left[-|\overline{V}|^2, |\overline{V}|^2]\right]$ and furthermore $\theta_{ij} \in \left[\underline{\Delta \theta},\overline{\Delta \theta}\right]$. We can narrow the bounds for $c_{ij}$ further to $|\underline{c_{ij}},\overline{c_{ij}}| = |\overline{V}|^2 cos(\underline{\Delta \theta}) \leq c_{ij} \leq |\overline{V}|^2$ ($\because 0 \in \left[\underline{\Delta \theta},\overline{\Delta \theta}\right]$). Thus, we can construct the McCormick envelopes by replacing the bilinear term with an auxiliary variable $w_{ij} = c_{ij}\theta_{ij}$ with the following associated constraints:

\begin{align}
& w_{ij} \geq \underline{c_{ij}} \theta_{ij} + c_{ij} \underline{\theta_{ij}}-\underline{c_{ij}} \underline{\theta_{ij}} \label{eq:mce1}\\
& w_{ij} \geq \overline{c_{ij}} \theta_{ij}+c_{ij} \overline{\theta_{ij}}-\overline{c_{ij}} \overline{\theta_{ij}} \label{eq:mce2} \\
& w_{ij} \leq \overline{c_{ij}} \theta_{ij}+c_{ij} \underline{\theta_{ij}}-\overline{c_{ij}} \underline{\theta_{ij}} \label{eq:mce3}\\
& w_{ij} \leq c_{ij} \overline{\theta_{ij}}+\underline{c_{ij}} \theta_{ij}-\underline{c_{ij}} \overline{\theta_{ij}}\label{eq:mce4}
\end{align}

% which for our case reduces to:
% \begin{align}
% & w_{ij} \geq -|\overline{V}|^2 \theta_{ij} + c_{ij} \underline{\Delta \theta} + |\overline{V}|^2 \underline{\Delta \theta} \\
% & w_{ij} \geq |\overline{V}|^2 \theta_{ij}+c_{ij} \overline{\Delta \theta}-|\overline{V}|^2 \overline{\Delta \theta} \\
% & w_{ij} \leq |\overline{V}|^2 \theta_{ij}+c_{ij} \underline{\Delta \theta}-|\overline{V}|^2 \underline{\Delta \theta} \\
% & w_{ij} \leq c_{ij} \overline{\Delta \theta}-|\overline{V}|^2 \theta_{ij} + |\overline{V}|^2 \overline{\Delta \theta}
% \end{align}

Of course, the tightness of this MCE relaxation depends heavily on the variable bounds used to construct the convex underestimators and concave overestimators. The relaxation could be improved more by iteratively decreasing the upper bounds and increasing the lower bounds, through piecewise MCE relaxations or other similar approaches. However, I have not considered those yet here.

% We can actually obtain even tighter bounds on $c_{ij}$ by noting that $-1.047 = \underline{\Delta \theta} \leq \theta_{ij} \leq \overline{\Delta \theta} = 1.047 \; rad$, and $cos(1.047) \approx \frac{1}{2} \leq cos(\theta) \leq 1$ for $\theta \in [-1.047, 1.047]$. This implies that $\frac{|\overline{V}|^2}{2} \leq c_{ij} \leq |\overline{V}|^2$. Plugging these into \cref{eq:mce1,eq:mce2,eq:mce3,eq:mce4} gives us:

% \begin{align*}
%     & w_{ij} \geq \frac{|\overline{V}|^2 \theta_{ij}}{2} + c_{ij} \underline{\Delta \theta} -\frac{|\overline{V}|^2 \underline{\Delta \theta}}{2} \\
%     & w_{ij} \geq |\overline{V}|^2 \theta_{ij}+c_{ij} \overline{\Delta \theta}-|\overline{V}|^2 \overline{\Delta \theta} \\
%     & w_{ij} \leq |\overline{V}|^2 \theta_{ij}+c_{ij} \underline{\Delta \theta}-|\overline{V}|^2 \underline{\Delta \theta} \\
%     & w_{ij} \leq c_{ij} \overline{\Delta \theta}+\frac{|\overline{V}|^2 \theta_{ij}}{2} - \frac{|\overline{V}|^2 \overline{\Delta \theta}}{2}
% \end{align*}

% \subsubsection{Disjunctive or conditional constraints}

We introduce additional binary variables $a_{ij} \in \{0,1\}$ for each line $ij \in L$ that indicates whether a line is open or closed. In order to represent switching, we need to enforce that both the real and reactive power flows on an open line are constrained to zero, since an open circuit cannot transfer any current. We thus need to modify the constraints \cref{eq:kirchoff1,eq:kirchoff2} accordingly:
\begin{align}
    p_{i j}=a_{ij}\left(-G_{i j} e_i+G_{i j} c_{i j}+B_{i j} s_{i j}\right), \quad \forall(i, j) \in L \label{eq:kirchoff1_sw}\\
   q_{i j}=a_{ij}\left(\left(B_{i j}-b_{i j} / 2\right) e_i-B_{i j} c_{i j}+G_{i j} s_{i j}\right), \quad \forall(i, j) \in L \label{eq:kirchoff2_sw}
\end{align}
Since we don't know the exact bounds on the continuous variables (terms inside the parentheses), an exact linear reformulation of this bilinear constraint is not possible. However we can represent this using as the following set of disjunctive or conditional equalities:

\begin{align}
    \left[\begin{array}{c}
a_{ij} \\
p_{ij} = -G_{i j} e_i+G_{i j} c_{i j}+B_{i j} s_{i j} \\
q_{i j}=\left(B_{i j}-b_{i j} / 2\right) e_i-B_{i j} c_{i j}+G_{i j} s_{i j}
\end{array}\right] \vee\left[\begin{array}{c}
\neg a_{ij} \\
p_{ij} = 0\\
q_{ij}=0
\end{array}\right]
\label{eq:disjunctive_constr}
\end{align}

These can be relaxed to the following linear constraints using the big-M method, allowing us to get rid of the bilinear terms involving products of continuous and binary variables:
\begin{align*}
    - p_{ij} -G_{i j} e_i+G_{i j} c_{i j}+B_{i j} s_{i j} + (1 - a_{ij})M_{ij} & \geq 0 \\
    - p_{ij} -G_{i j} e_i+G_{i j} c_{i j}+B_{i j} s_{i j} - (1 - a_{ij})M_{ij} & \leq 0 \\
    - q_{i j} + \left(B_{i j}-b_{i j} / 2\right) e_i-B_{i j} c_{i j}+G_{i j} s_{i j} + (1 - a_{ij})N_{ij} & \geq 0 \\
    - q_{i j} + \left(B_{i j}-b_{i j} / 2\right) e_i-B_{i j} c_{i j}+G_{i j} s_{i j} - (1 - a_{ij})N_{ij} & \geq 0 \\
    -M_{ij} a_{ij} \leq p_{ij} \leq M_{ij} a_{ij}, \quad -M_{ij} a_{ij} \leq q_{ij} \leq M_{ij} a_{ij} & 
\end{align*}
where $M_{ij}, N_{ij}$ are chosen to be sufficiently large positive numbers - they are chosen such that $M_{ij} > \left|G_{i j} c_{i j}+B_{i j} s_{i j}-G_{i j} e_i\right|$ and $N_{ij} > \left|\left(B_{i j}-b_{i j} / 2\right) e_i-B_{i j} c_{i j}+G_{i j} s_{i j}\right|$.

% Do we need to also add the binary variables to the apparent line power flow thermal limit constraint too? i.e.
% $$
% \left(p_{i j}\right)^2+\left(q_{i j}\right)^2 \leq \left(\overline{f}_{ij}\right)^2, \quad \forall(i, j) \in L 
% $$

% What about Convex hull reformulation?

We need to ensure that only lines with either switches or circuit breakers can be made open. Suppose $SW$ is an $n \times n$ matrix that indicates whether a line has a switching device i.e., $SW_{ij} = 1$ and is $0$ otherwise. Then, the following constraint ensures that only valid switching actions are allowed:
\begin{align}
    1-a_{ij} \leq SW_{ij} \quad \forall (i,j) \in L
\end{align}
Finally, we can optionally add an upper limit on the total number of lines that can be opened in the optimized network:
$$
\sum_{(i,j)\in L} (1-a_{ij}) \leq N_{sw}
$$
This allows the grid operator to control exactly how many switching operations can be made at each timestep. This may be necessary for real systems where it may not be feasible to operate all available switches and circuit breakers. Thus, we have relaxed the original nonconvex mixed integer nonlinear program for OTS-ACOPF to a mixed integer second order conic convex program (MISOCP) - these are much more computationally tractable to solve than MINLPs.

\section{Numerical simulations and results}

I ran some preliminary simulations of the proposed formulations on a few different IEEE standard transmission test cases. These are relatively small networks but they still demonstrate some interesting results that I summarize below. Some notes on the simulations themselves:
\begin{itemize}
    \item For the larger systems, there were several cases where the solver had difficulties converging to the optimal solution or reported local infeasibility. This was especially a challenge for the MINLP formulations and/or while using free solvers like IPOPT or SCIP. \item In order to deal with this, I warm-started the optimization process by initializing the variables using solutions from DCOPF, more relaxed versions of the same problem or by temporarily removing certain constraints like the line flow limits. This allowed me to achieve convergence for all the cases.
    \item For simplicity, I used a fixed large value of $M=100$ for the big-M relaxations. This worked well since I already know most of my variables and terms are of the order of $\approx 1.0$ after per-unitization. However, tuning the value of $M$ for different cases could improve the performance and also avoid potential issues around ill-conditioning and scaling (although I didn't run into this during my simulations).
\end{itemize}

\subsection{Comparisons of different formulations}

I first compared the different formulations in terms of their computational performance, as shown in the table below. The formulations are defined as follows:
\begin{enumerate}
    \item DC OPF: The simplest version of the problem, convex quadratic program (with linear constraints).
    \item Non convex, nonlinear AC OPF: Exactly describes the power physics.
    \item SOCP relaxed OPF: Using second-order conic programming convex relaxation but still containing the nonconvex trigonometric voltage angle constraint.
    \item SOCP relaxed OPF + MCEs: Also relax the nonconvex trigonometric voltage angle constraints using a small angle approximation and McCormick envelopes (MCE).
    \item (1) with OTS
    \item (2) with OTS
    \item (3) with OTS, containing bilinear terms involving products of binary and continuous variables.
    \item (7) with big-M reformulations to remove the bilinear terms
    \item MISOCP by combining (8) with MCE relaxations of the angle constraint
\end{enumerate}

As expected intuitively, we see that the optimality gap in \cref{tab:comp_compare2} increases as we further relax and approximate the problem. Here I used the relative difference between optimal solution values to compare the gap between different formulations. The DC OPF (1) is significantly worse than all the other formulations, indicating that solutions from the DC OPF may not necessarily be optimal for the real system or may not even be AC feasible since it doesn't take into account all the constraints. The other relaxations work reasonably well although their exact performance depends on the specific test case being considered. For the nominal operation, the SOCP relaxed OPF (3) works best and further convexifying the angle constraint via MCEs in (4) only marginally increases the optimality gap. This suggests that the MCE approach is valid at least for these test cases. In terms of the OTS formulations, we see that final convex MISOCP formulation with MCEs (9) still has a relatively small optimality gap. One interesting point is that the big-M reformulation in (8) performs much better than (7) on the 9-bus case but worse on the 39-bus case. This is something I still need to investigate further. 

DC OPF is significantly faster than the other methods in most cases, owing to its simple model but this comes at the cost of lower accuracy and we may need to check for AC feasibility of the obtained solutions Interestingly, we find that some of the relaxations actually take slightly longer than their nonconvex, stricter counterparts. This may be due to the additional variables we needed to introduce to deal with the nonlinearities, which increase the dimensionality. However, I expect that the relaxed formulations may scale better for larger networks - I still need to test this hypothesis through more simulations going forward. Another encouraging trend from \cref{tab:comp_compare1} is that the relaxations seem to offer more of an incremental benefit (in terms of faster solution times) when applied to the OTS problem to solve the MINLPs, as opposed to the nominal case without OTS where we're solving a simpler problem. 

While I observed low optimality gaps with the relaxations for these experiments, this is not guaranteed for all situations. Further refinements can be made to restrict the feasible space, obtain tighter relaxations and ensure that the obtained solutions are feasible for the original nonconvex program. Nevertheless, applying any of the above ACOPF-based relaxations would still offer better approximations than the crude DCOPF approximation.

% Local vs globally optimal solution guarantees. Minimal optimality gap
% Checking AC feasibility of solutions from the relaxed problem?
% All the methods and relaxations returned very similar optimal solution values across all three test cases, indicating minimal optimality gap at least for these relatively small systems. 
% This motivate the use of such relaxation

\begin{table}[H]
\centering
\begin{tabular}{@{}lcc@{}}
\toprule
                                                & \multicolumn{2}{c}{Runtime (s)}                    \\ \midrule
\textbf{Different formulations}                 & \textbf{9-bus} & \textbf{39-bus} \\ \midrule
(1) DC OPF                                          &   0.0033    &  0.0061\\
(2) Non convex AC OPF                               &  0.0249      &  0.744    \\
(3) SOCP relaxed OPF                                &  0.045      &  0.639    \\
(4) SOCP relaxed OPF + MCEs                         &  0.0556     &   0.677      \\ 
\midrule
\multicolumn{3}{c}{\textbf{Simulations run with no constraints switching actions} $N_{sw}$}          \\ 
\midrule
(5) DC OPF OTS                                      &   0.0575    &     0.555         \\
(6) Non convex ACOPF OTS                            &   0.0806      &    0.376             \\
(7) OTS with SOCP relaxation + bilinear terms       &    0.0728     &     0.402              \\
(8) OTS with SOCP relaxation + Big-M reformulations &    0.0437     &      0.387           \\
(9) MISOCP relaxed OTS with Big-M + MCEs            &     0.0815     &     0.527                  \\ \bottomrule
\end{tabular}
\caption{\label{tab:comp_compare1} Computational comparison of different OPF and OTS approaches - for a fair comparison these were all solved using the free, open-source SCIP solver in Julia.}
\end{table}

\begin{table}[H]
\centering
\begin{tabular}{@{}lcc@{}}
\toprule
                                                & \multicolumn{2}{c}{Optimality gap (\%)}                    \\ \midrule
\textbf{Different formulations}                 & \textbf{9-bus} & \textbf{39-bus} \\ \midrule
(1) DC OPF                                          &   0.0891     &    1.091    \\
(2) Non convex AC OPF (exact)                           &  0       &     0    \\
(3) SOCP relaxed OPF                                &   3.712e-10     &  1.955e-14    \\
(4) SOCP relaxed OPF + MCEs                         &   1.115e-9   &    4.671e-12   \\ \midrule
\multicolumn{3}{c}{\textbf{Simulations run with no constraints switching actions} $N_{sw}$}          \\ \midrule
(5) DC OPF OTS                                      &    0.0891      &     0.024        \\
(6) Non convex ACOPF OTS                            &    0      &   0          \\
(7) OTS with SOCP relaxation + bilinear terms       &   7.71e-7     &     1.87e-14            \\
(8) OTS with SOCP relaxation + Big-M reformulations &   1.639e-14     &   2.307e-12              \\
(9) MISOCP relaxed OTS with Big-M + MCEs            &   6.988e-8      &   8.906e-10             \\ \bottomrule
\end{tabular}
\caption{\label{tab:comp_compare2} Optimality comparison of different OPF and OTS approaches.}
\end{table}

\subsection{Effects of OTS on the system}

\begin{figure}[H]
\centering
     \includegraphics[width=\columnwidth]{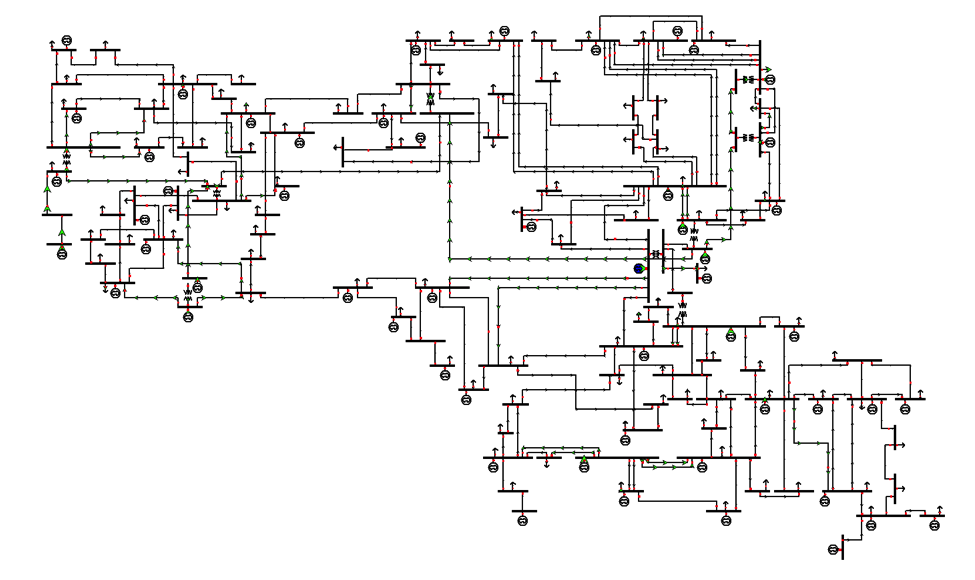}
     \caption{IEEE-118 bus system. \label{fig:118}}
\end{figure} 

Although I tested all the nine formulations on the test cases, for conciseness here I'm only reporting the results for (i) the nominal case while applying approach (4) vs (ii) the OTS case by applying (9) on the larger IEEE-118 bus system. I simulated the OPF dispatch using load and generation data for 1 hour. This is because I did not observe significant impacts of OTS on the smaller 9-bus and 39-bus systems, indicating that they are not really transmission capacity constrained. The IEEE-118 bus system consists of 19 generators, 35 synchronous condensers, 177 lines, 9 transformers, and 91 loads, shown in \cref{fig:118}. These simulations were run using the Gurobi solver in Julia. We assess the impacts of OTS by examining some important metrics. We can obtain the locational marginal prices (LMP) at each bus as the dual multipliers associated with the real power (P) balance constraints. These are how electricity prices are set in the wholesale market as well. Difference in LMPs across different buses in the network indicate that there is congestion, i.e. the network operation is constrained by the transmission capacity of the lines (power flow limits). The effect of congestion on system performance can be quantified via the congestion rent:
$$
\text {Congestion rent} = CR = \left(\sum_{\mathrm{i}=1}^{n_D} p^i_{D} \cdot L M P_i\right)-\left(\sum_{j=1}^{n_G} p^G_j . L M P_j\right)
$$
This is the difference between the total amount paid by electricity consumers minus the costs of procuring it from the generators, thus $CR > 0$ indicates that there are inefficiencies in the market allocation due to the effects of congestion and losses. Another key metric is the total system-wide cost of meeting the demand, given by the total generation cost. This can be directly queried as the objective function value at the optimal solution:
$$
\sum_{i \in G} f_i\left(P_{G i}^*\right) = \sum_{i \in G} C^{G1}_i \left(p^G_i\right)^2 + C^{G2}_i p^G_i + C^{G13}_i
$$
where we used a quadratic variable cost function which is characteristic of conventional fossil fuel based generators. Finally, we can quantify the extent to which transmission capacity is constrained through the flowrate marginal prices (FMPs) which correspond to the dual variables of the line flow limit inequality constraints \cref{eq:flow_lim_socp}. These are zero for lines that are not congested since the constraint with strict inequality, but screening for non-zero values allows us to identify which lines are congested and to what extent.

% For most of the simulations, the LMPs were actually very similar across all the nodes in the network. This indicates that there is little to no congestion and there is sufficient transmission capacity to serve all the load without running into issues.

% Don't see much impacts on IEEE 9, 30 and 39-bus systems since these are relatively simple systems with small numbers of branches. Also show table with average (and maximum) flow marginal prices and SD in LMPs to show that there isn't much congestion to begin with. The system isn't really transmission capacity constrained and has enough slack to accommodate changes in generation profiles.

% Optimal generator dispatch still changes with OTS? 

% Compare nominal case solved as SOCP with MCE relaxations vs its MISOCP OTS counterpart, both solved using Gurobi.

\begin{table}[H]
\centering
\begin{tabular}{@{}lcc@{}}
\toprule
                                      & \textbf{Nominal} & \textbf{OTS} \\ \midrule
\multicolumn{3}{c}{\textbf{With no limit on $N_{sw}$}}                                                      \\ \midrule
\textbf{Total generation cost (\$/h)} &     62,165   &   60,018  (-3.45\%)                 \\
\textbf{Congestion rent (\$/h)}       &    39,433      &     2579 (-93.45\%)       
\\
\textbf{Average LMP (\$/MWh)}         &     44.23       &      33.84 (-23.49\%)                \\
\textbf{Average |FMP| (\$/MWh)}         &   0.7404         &      0.0176  (-97.62\%)                \\
\textbf{LMP standard deviation (\$/MWh)}      &    11.89      &      0.775 (-93.49 \%)                \\
\textbf{No. of congested lines}      &        5                    &  4 \\
\textbf{Optimal no. of open lines}      &         N/A                   & 32 (out of 186 total) \\
\bottomrule 
\end{tabular}
\caption{Effects of OTS on the IEEE-118 bus test system \label{tab:compare3}.}
\end{table}

We find that applying OTS creates significant benefits especially in terms of reducing system-wide congestion. This results in lower LMPs and FMPs, and thus lower congestion rents. Improved dispatch allow us to shift some generation to cheaper sources, lowering total costs slightly as well. The lower spatial volatility in LMPs also indicates lower congestion and more efficient network operation. For simplicity, these simulations were run with no constraint $N_{sw}$. Thus, it's important to note that these benefits may be less pronounced if there is an upper limit imposed by the operator on the number of lines allowed to be open or turned `OFF' ($N_{sw}$).

\section{Conclusions and future work}

In this study, I formulated and simulated various different optimal power flow relaxations and applied them to the optimal transmission switching problem. I built upon previous work to include more accurate models and constraints while performing OTS. In addition to comparing the runtime and performance of these approaches on 3 different realistic test networks, I also demonstrated the benefits of applying OTS on a larger system. It helps reduce overall costs, reduce congestion and result in more efficient power dispatch system-wide. For future work, I plan to scale up this approach to much larger transmission grids in order to more rigorously compare the different formulations. This will also allow me to assess what the true potential of OTS is. 

\bibliographystyle{IEEEtran}
\bibliography{references}

\end{document}